\documentclass[12pt,english,floatfix,preprint,amsmath,amssymb,aps,prd,superscriptaddress,titlepage]{revtex4-2}

\usepackage{graphicx}
\usepackage[utf8]{inputenc}
\usepackage[T1]{fontenc} 
\usepackage{microtype} 
\usepackage{physics}    
\usepackage{tensor}
\usepackage[svgnames]{xcolor} 
\usepackage[obeyFinal, 
    color       = LightGray,
    bordercolor = LightGray,
    textsize    = footnotesize,
    figwidth    = 0.99\linewidth,
    prependcaption]{todonotes} 
\usepackage[allcolors=blue,colorlinks,]{hyperref} 
\usepackage{orcidlink} 

\begin{document}

\title{Bounds on Black Bounces from Phenomenological Imprints}
\author{Alana C. L. Santos\,{\orcidlink{0000-0002-0080-5699}}}
\email{alanasantos@fisica.ufc.br}
\affiliation{Universidade Federal do Ceará (UFC), Departamento de Física,
 Campus do Pici, Fortaleza- CE, C.P. 6030, 60455-760- Brazil}
 \author{Marcos S. Melo\,{\orcidlink{}}}
\email{marcos.melo@fisica.ufc.br }
\affiliation{Universidade Federal do Ceará (UFC), Departamento de Física,
 Campus do Pici, Fortaleza- CE, C.P. 6030, 60455-760- Brazil}
\author{Roberto V. Maluf\,{\orcidlink{0000-0002-9952-4589}}}
\email{r.v.maluf@fisica.ufc.br}
\affiliation{Universidade Federal do Ceará (UFC), Departamento de Física,
 Campus do Pici, Fortaleza- CE, C.P. 6030, 60455-760- Brazil}

\date{\today}

\begin{abstract}
In this work, we derive phenomenological constraints on a symmetric black-bounce geometry supported by an anisotropic fluid. We investigate the effects of the bounce parameter and the matter parameter on classical and strong-field observables, considering the perihelion advance, light deflection, Shapiro time delay, and black-hole shadow. In the weak-field regime, we obtain analytical corrections to the corresponding general-relativistic predictions and use Solar-System measurements to constrain the parameter space. We find that the matter parameter is generally constrained more strongly than the minimum-radius scale, reflecting the suppressed dependence of weak-field observables on the bounce parameter. In the strong-field regime, we show that the asymptotic shadow radius depends directly on the matter parameter, whereas the bounce parameter enters through the existence condition of the outer unstable photon orbit rather than through an independent correction to the shadow size. These results highlight the distinct roles played by the matter and bounce parameters in weak- and strong-field observables and provide complementary constraints on the parameter space of the symmetric black-bounce geometry.

\end{abstract}

\maketitle

\section{Introduction}\label{sec: intro}
General Relativity (GR) provides the most experimentally successful description of gravitation over the scales at which it has been tested. Nevertheless, under physically reasonable assumptions, the theory may admit geodesically incomplete spacetimes, motivating the search for effective geometries in which the central singularity is regularized \cite{Will:2014kxa}. A prominent example is provided by black-bounce spacetimes, characterized by a nonvanishing minimum areal radius that replaces the singular center and allows for configurations describing regular black holes, wormholes or extremal objects \cite{Simpson:2018tsi}. This has motivated considerable effort to identify the matter sources capable of supporting such geometries, within GR and beyond \cite{Junior:2026ism,B:2026jtm,Lobo:2026flk,Alencar:2026qeb,Cordeiro:2025ivw,Ling:2025ncw,Bronnikov:2022bud,Canate:2022gpy}. Since black-bounce geometries can closely mimic several phenomenological features of conventional black holes, considerable effort has been devoted to identifying observational signatures capable of distinguishing them from their singular GR counterparts \cite{Dasgupta:2025qwq, Martinez-Guerrero:2026unh, Silva:2026mlo, Siqueira:2026uzu}. In particular, their phenomenology has been investigated through optical observables \cite{Guerrero:2021ues,Olmo:2023lil,Ovejero-Bermudez:2026jja,Ahmed:2026bwm,Nascimento:2025mtr,Bambhaniya:2021ugr,Guo:2021wid}, superradiant phenomena \cite{Franzin:2022iai}, gravitational lensing \cite{Islam:2021ful,He:2024ozb,Nascimento:2020ime,Cheng:2021hoc,Tsukamoto:2021caq}, and quasinormal-mode spectra \cite{Duran-Cabaces:2025sly,Yang:2021cvh,Yang:2026dyonic}.

The relation between the matter content and the radial metric structure of black-bounce was investigated in Ref.~\cite{Lessa:2024erf}. The authors demonstrated that black-bounce geometries can be consistently supported by an anisotropic-fluid without imposing a \emph{priori} the usual condition $g_{tt}=g_{rr}^{-1}$ in a general static and spherically symmetric spacetime. In particular, they obtained a symmetric solution that preserves the areal-radius of the Simpson--Visser geometry \cite{Simpson:2018tsi}, while treating the temporal and radial metric functions as independent functions to be determined self-consistently from the Einstein field equations. This additional freedom in the metric structure leads to distinctive features in both the perturbative and optical sectors, such as novel signatures in gravitational-wave echoes \cite{Santos:2025xbk} and multiple photon-ring structures \cite{Santos:2026}. Such a framework therefore provides a natural setting for investigating how relaxing the Schwarzschild-like relation between the metric components affects observable properties of black-bounce spacetimes.

A natural question that emerges from an observational perspective is how large the minimum areal radius can be while remaining compatible with current experimental constraints. Classical gravitational tests provide a particularly useful way of addressing this question, since small deviations from the GR predictions can be translated into bounds on the parameters characterizing the underlying geometry \cite{Junior:2023nku}. Similar strategies have been widely employed in modified-gravity scenarios, for instance to constrain Lorentz-violating parameters through Solar-System observations \cite{Casana:2017jkc,Bailey:2009me,Tso:2011up,Belchior:2025xam,AraujoFilho:2024ykw,Filho:2022yrk,Yang:2023wtu}. Similar constraints have also been derived for the Simpson--Visser geometry \cite{Ovejero-Bermudez:2026jja,Zhou:2020zys} and its charged extension \cite{Zhang:2022zox,Zhang:2022nnj}. Beyond the weak-field regime, constraints have also been investigated using strong-field observations, particularly those associated with the shadows of Sgr $A^{*}$ and $M87^{*}$. The basic procedure consists of comparing the theoretically predicted shadow size with the observational bounds inferred from the measured angular size and the corresponding mass-to-distance ratio, typically within the $1\sigma$ and $2\sigma$ confidence intervals \cite{Vagnozzi:2022moj,EventHorizonTelescope:2021dqv,Jafarzade:2021umv}. It is therefore relevant to investigate how the phenomenology is modified when the Schwarzschild-like relation between the temporal and radial metric functions is relaxed and, more importantly, whether current observations can place quantitative constraints on the minimum-radius parameter of the resulting geometry.

Motivated by this question, in this work we investigate observational constraints on the black-bounce parameter $r_{0}$ across different gravitational regimes. At Solar-System scales, we consider three classical tests: the advance of the perihelion, the deflection of light and the time delay. We then extend the analysis to the strong-field regime by studying black-hole-shadow observables, in particular the shadow radius and the corresponding mass-to-distance angular scale. By comparing the theoretical predictions with observational uncertainties,
we derive constraints on the model parameters and investigate whether the different observables can independently constrain the minimum-radius scale $r_0$. 

\section{Symmetric Black-Bounce Model}\label{secII}
First, we briefly review the construction of the symmetric black-bounce geometry. Let us consider the general static and spherically symmetric line element,
\begin{equation}\label{r}
ds^2=-A(r)dt^2+B(r)dr^2+\Sigma(r)^2d\Omega^2,
\end{equation}
where $\Sigma(r)$ denotes the areal radius. A black-bounce geometry is characterized by a nonvanishing minimum of $\Sigma(r)$ satisfying
\begin{equation}
\Sigma(r_0)\neq 0, \quad   \Sigma'(r_0) =0 , \quad \Sigma''(r_0) >0, 
\end{equation}
where $r_0$ defines the minimum areal radius, whereas event horizons are independently determined by the zeros of $A(r)$. As shown in Ref.~\cite{Lessa:2024erf}, symmetric and asymmetric black-bounce solutions can be generated within GR considering an anisotropic fluid, $T^\mu{}_\nu=\mathrm{diag}(-\rho(r),p_r(r),p_t(r),p_t(r))$,
where $\rho(r)$, $p_r(r)$, and $p_t(r)$ denote the energy density, radial pressure, and tangential pressure, respectively, satisfying the equations of state
$\rho(r)+p_r(r)=0$ and $p_t(r)=\omega\rho(r)$. The conservation equation and the Einstein field equations then yield
\begin{equation}
\Sigma(r)=\frac{\Sigma_0}{\rho(r)^{\frac{1}{2(\omega+1)}}}, \quad B(r)=\frac{\Sigma'(r)^2}{A(r)}.
\end{equation}
This also allows us to adopt $\Sigma$ as the radial coordinate and write
\begin{equation}\label{sigma}
ds^2=-A(\Sigma)dt^2+\frac{d\Sigma^2}{A(\Sigma)}+\Sigma^2d\Omega^2.
\end{equation}
It is important to note that this representation is restricted to regions where $\Sigma'(r)\neq0$ and therefore does not explicitly retain the information associated with the minimum of the areal radius. For the symmetric solution of Ref.~\cite{Lessa:2024erf}, the energy density is chosen as
\begin{equation}
\rho(r)=\frac{\rho_0}{(r_0^2+r^2)^{1+\omega}},
\end{equation}
which yields the areal function
\begin{equation}
\Sigma(r)=\sqrt{r_0^2+r^2},
\end{equation}
and, together with the Einstein field equations, leads to the line element
\begin{equation}
ds^2=-A(r)dt^2+ \frac{dr^2}{\left(1+\frac{r_0^2}{r^2}\right)A(r)}+(r_0^2+r^2)d\Omega^2,
\end{equation}
where,
\begin{equation}
A(r) = 1 - \frac{2M}{\sqrt{r_0^2 + r^2}} + \frac{\rho_0}{(2\omega - 1)(r_0^2 + r^2)^\omega}.    
\end{equation}
$r_0$ sets the minimum areal radius and controls the regularization of the central region. In the limit $r_0\rightarrow0$, the Kiselev geometry is recovered \cite{Kiselev:2002dx}. In particular, for $\omega=1$,
\begin{equation}
A(\Sigma)=1-\frac{2M}{\Sigma}+\frac{\rho_0}{\Sigma^2},
\end{equation}
which has the same functional form as the Reissner--Nordstr\"om metric, with $\rho_0$ playing the role of an effective charge parameter. Depending on the values of $r_0$, $M$ and $\rho_0$, the resulting spacetime may describe a regular black hole, an extremal configuration or a traversable wormhole.

\section{CLASSICAL TESTS}\label{secIII}
Since our main goal is to extract analytical information from these classical tests, we first adopt the coordinate system in Eq.~(\ref{sigma}), as has already been done for the Simpson--Visser case \cite{Ovejero-Bermudez:2026jja,Nascimento:2020ime, Zhang:2022zox, Zhang:2022nnj}. This choice allows us to formulate the problem in close analogy with the Reissner--Nordstrom case. After obtaining the corresponding analytical expressions, we transform the results back to the original coordinate system defined in Eq.~(\ref{r}), in which the dependence on the characteristic radius $r_{0}$ becomes explicit. In the remainder of this work, we restrict the analysis to the $\omega=1$ sector.

\subsection{ADVANCE OF PERIHELION}
We begin by considering the motion of massive test particles. Given the static and spherically symmetric nature of the spacetime, the motion can be restricted, without loss of generality, to the equatorial plane, $\theta = \frac{\pi}{2}$. The Killing vectors associated with time translations and axial rotations, $\partial_t$ and $\partial_\phi$, respectively, imply the conservation of energy and angular momentum.
\begin{equation}
E=A(\Sigma)\dot{t}, \qquad L=\Sigma^{2}\dot{\phi}.    
\end{equation}
From the test-particle Lagrangian, we obtain the radial geodesic equation
\begin{equation}
 \dot{\Sigma}^{2} = E^{2} - A(\Sigma) \left(1+\frac{L^{2}}{\Sigma^{2}}\right),    
\end{equation}
where the dot represents derivative with respect to an affine parameter denoted by $\tau$. Introducing $U = \frac{1}{\Sigma}$ and using $\Sigma = \Sigma(\phi)$, the radial equation can be rewritten in the form of the Binet equation
\begin{equation}\label{binet}
\frac{d^{2}U}{d\phi^{2}} +U = \frac{M}{L^{2}} +3MU^{2} -\frac{\rho_{0}}{L^{2}}U -2\rho_{0}U^{3}.   
\end{equation}
To make the perturbative ordering explicit, we introduce the parameters $\lambda$, $\epsilon$ and $q$ as
\begin{equation}
\lambda \equiv \frac{L^2U}{M},\quad \epsilon \equiv \frac{3M^{2}}{L^{2}},\qquad q \equiv \frac{\rho_{0}}{L^2},
\end{equation}
where $\epsilon\ll1, |q|\ll1$. In terms of these variables, Eq.~(\ref{binet}) becomes
\begin{equation}\label{binetlambda}
 \lambda''+\lambda = 1 +\epsilon\lambda^{2} -q\lambda -\frac{2}{3}\epsilon q\,\lambda^{3},    
\end{equation}
where a prime denotes differentiation with respect to $\phi$. To solve this equation, we employ the Homotopy Perturbation Method (HPM) \cite{Shchigolev:2015sgg}, assuming that the solution can be expanded as a power series
\begin{equation}
\lambda = \lambda_{0} + \lambda_{1} + ...
\end{equation}
At zeroth order in $\epsilon$ and $q$, Eq.~(\ref{binetlambda}) reduces to
\begin{equation}
\lambda_{0}''+\lambda_{0}=1.
\end{equation}
The corresponding solution describes the Newtonian Keplerian orbit and can be written as
\begin{equation}
\lambda_{0}(\phi) = 1+e_{\Sigma}\cos\phi,
\end{equation}
where $e_{\Sigma}$ denotes the orbital eccentricity associated with the areal coordinate $\Sigma$. Equivalently, the zeroth-order solution for $U$ is
\begin{equation}\label{zeroorder}
U_{0}(\phi) = \frac{M}{L^2}\left(1+e_{\Sigma}\cos\phi\right).
\end{equation}
On the other hand, the first-order solution must satisfy
\begin{equation}\label{lambda}
\lambda_{1}''+\lambda_{1} = \epsilon \left(1+\frac{e_{\Sigma}^{2}}{2}\right) -q + e_{\Sigma}(2\epsilon-q)\cos\phi + \frac{\epsilon e_{\Sigma}^{2}}{2}\cos2\phi .    
\end{equation}
The constant contribution produces only a displacement of the orbit, whereas the $\cos2\phi$ contribution represents a bounded oscillatory correction. The term proportional to $\cos\phi$, on the other hand, is resonant with the homogeneous solution and generates a secular contribution responsible for the perihelion advance. A particular solution of Eq. (\ref{lambda}) is
\begin{equation}
 \lambda_{1} = \epsilon\left(1+\frac{e_{\Sigma}^{2}}{2}\right) -q - \frac{\epsilon e_{\Sigma}^{2}}{6}\cos2\phi + \frac{e_{\Sigma}}{2}(2\epsilon-q)\phi\sin\phi,   
\end{equation}
where homogeneous contributions proportional to $\cos\phi$ and
$\sin\phi$ have been absorbed into a redefinition of the orbital eccentricity and initial phase. Therefore, up to first perturbative order, the orbital solution can be written as
\begin{equation}
\lambda(\phi) \simeq 1+e_{\Sigma}\cos\phi + \epsilon \left( 1+\frac{e_{\Sigma}^{2}}{2} \right) -q - \frac{\epsilon e_{\Sigma}^{2}}{6}\cos2\phi + \frac{e_{\Sigma}}{2}(2\epsilon-q)\phi\sin\phi .    
\end{equation}
Noting that, to first order in $\epsilon$ and $q$,
\begin{equation}
\cos\left[\left(1-\epsilon+\frac{q}{2}\right)\phi\right] \simeq \cos\phi +\left(\epsilon-\frac{q}{2}\right)\phi\sin\phi, \qquad \left|\epsilon-\frac{q}{2}\right|\ll1,
\end{equation}
the perturbative solution can be recast as
\begin{equation}
\lambda(\phi) \simeq 1+\epsilon\left(1+\frac{e_{\Sigma}^{2}}{2}\right)-q -\frac{\epsilon e_{\Sigma}^{2}}{6}\cos 2\phi +e_{\Sigma} \cos\left[\left(1-\epsilon+\frac{q}{2}\right)\phi\right].
\end{equation}
The secular contribution can therefore be interpreted as a small shift in the orbital frequency. A complete radial oscillation is obtained when
\begin{equation}
\left(1-\epsilon+\frac{q}{2}\right)\phi=2\pi.
\end{equation}
Hence, the azimuthal angle accumulated between two successive perihelia is
\begin{equation}
\phi = \frac{2\pi}{1-\epsilon+q/2},
\end{equation}
and the corresponding perihelion advance per radial period is
\begin{equation}
\Delta\phi \equiv \phi-2\pi \simeq
2\pi\left(\epsilon-\frac{q}{2}\right) + \mathcal{O}(\epsilon^{2},q^{2},\epsilon q).
\end{equation}
In terms of the conserved quantities, the perihelion advance is given by
\begin{equation}
\Delta\phi = \frac{6\pi M^2}{L^2} - \frac{\pi\rho_0}{L^2} + \mathcal{O}\left(\epsilon^{2},q^{2},\epsilon q\right). \label{eq:perihelion_EL}
\end{equation}
This result has the same form as in the
Reissner--Nordstr\"om geometry \cite{Chakraborty:2012sd,Hu:2013eya}.
At the Newtonian level, or zeroth-order, the angular momentum is related to the orbital elements associated with the areal radial coordinate $\Sigma$ by
\begin{equation}\label{eq:L2_areal}
L^2 = M a_\Sigma \left(1-e_\Sigma^2\right),
\end{equation}
where $a_\Sigma$ and $e_\Sigma$ denote the semi-major axis and eccentricity of the orbit parametrized in terms of the areal radius. Hence,
\begin{equation}\label{eq:perihelion_sigma}
\Delta\phi = \frac{\pi\left(6M^2-\rho_0\right)}
{M a_\Sigma\left(1-e_\Sigma^2\right)} + \mathcal{O}\left(\epsilon^{2},q^{2},\epsilon q\right).
\end{equation}
Although the use of the areal radius considerably simplifies the geodesic equations, the original radial coordinate $r$ is more suitable for describing the global black-bounce structure. We therefore reexpress the RN-like result
(\ref{eq:perihelion_sigma}) in terms of orbital elements defined with respect to the original radial coordinate $r$. We introduce the radial turning points as
\begin{equation}\label{eq:r_turning_points}
r_p=a_r(1-e_r), \qquad r_a=a_r(1+e_r),
\end{equation}
where $a_r$ and $e_r$ denote, respectively, the semi-major axis and eccentricity associated with the $r$-coordinate parametrization. The corresponding areal radii at the pericenter and apocenter are then
\begin{equation}\label{eq:sigma_turning_points}
\Sigma_p = \sqrt{a_r^2(1-e_r)^2+r_0^2}, \qquad \Sigma_a = \sqrt{a_r^2(1+e_r)^2+r_0^2}.
\end{equation}
On the other hand, the same turning points can be written in terms of the orbital elements associated with the areal coordinate as
\begin{equation}
\Sigma_p=a_\Sigma(1-e_\Sigma), \qquad \Sigma_a=a_\Sigma(1+e_\Sigma).
\end{equation}
Therefore,
\begin{equation}\label{eq:inverse_pSigma}
\frac{1} {a_\Sigma(1-e_\Sigma^2)} = \frac{1}{2}
\left(\frac{1}{\Sigma_p} + \frac{1}{\Sigma_a}\right).
\end{equation}
Substituting Eq.~(\ref{eq:sigma_turning_points}) into
Eq.~(\ref{eq:inverse_pSigma}), the perihelion advance can be
reexpressed directly in terms of the orbital parameters associated with the original radial coordinate as
\begin{equation}\label{eq:perihelion_r_exact}
\Delta\phi = \frac{\pi(6M^2-\rho_0)}{2M}\Bigg[
\frac{1}{\sqrt{a_r^2(1-e_r)^2+r_0^2}} + \frac{1}{\sqrt{a_r^2(1+e_r)^2+r_0^2}}\Bigg] + \mathcal{O}\left(\epsilon^{2},q^{2},\epsilon q\right).
\end{equation}
To obtain an analytical weak-field expression, we further assume that
the bounce scale is small compared with the orbital turning points, namely,
\begin{equation}\label{eq:bounce_orbital_expansion_condition}
\frac{r_0^2}{a_r^2(1-e_r)^2}\ll 1, \qquad \frac{r_0^2}{a_r^2(1+e_r)^2}\ll 1.
\end{equation}
The first condition is the more restrictive one, since the pericenter corresponds to the smallest radial distance along the orbit. Eq.~(\ref{eq:perihelion_r_exact}) gives
\begin{equation}\label{eq:perihelion_r_expanded}
\Delta\phi \simeq \frac{6\pi M} {a_r(1-e_r^2)} -
\frac{\pi\rho_0}{M a_r(1-e_r^2)} - \frac{3\pi M r_0^2(1+3e_r^2)}{ a_r^3(1-e_r^2)^3}
+ \frac{\pi\rho_0 r_0^2(1+3e_r^2)}{2M a_r^3(1-e_r^2)^3} +
\mathcal{O}\left(r_0^4,\epsilon^{2},q^{2},\epsilon q\right).
\end{equation}
The first term corresponds to the usual Schwarzschild result. Assuming a positive energy density, compatibility with the observational uncertainty $\sigma_\phi$ then requires
\begin{equation}
\Delta\phi-\Delta\phi_{\rm Schw} =\left|-\frac{\pi\rho_0}{M a_r(1-e_r^2)}+\frac{\pi r_0^2(1+3e_r^2)}{2M a_r^3(1-e_r^2)^3}\left(\rho_0-6M^2\right)
\right|\leq\sigma_\phi .
\label{eq:perihelion_constraint}
\end{equation}
For $0<\rho_0<6M^2$ both contributions to $\Delta\phi-\Delta\phi_{\rm Schw}$ have the same sign, and therefore
no cancellation occurs. In this regime, Eq.~(\ref{eq:perihelion_constraint}) gives
\begin{equation}
r_{0,\rm max}(\rho_0) = \left[\frac{2M a_r^3(1-e_r^2)^3}{\pi(1+3e_r^2)(6M^2-\rho_0)}\left(\sigma_\phi-\frac{\pi\rho_0}
{M a_r(1-e_r^2)}\right)\right]^{1/2}.
\label{eq:r0max_perihelion}
\end{equation}
For this upper bound to be real, one must also require
\begin{equation}
\rho_0 \leq\frac{\sigma_\phi M a_r(1-e_r^2)}{\pi}.
\label{eq:rho_perihelion_bound}
\end{equation}
At $\rho_0=6M^2$, the contribution proportional to $r_0^2$ vanishes at this perturbative order. For $\rho_0>6M^2$, the $\rho_0$ and $r_0^2$ contributions enter with opposite signs and may partially cancel.

\subsection{BENDING OF LIGHT}
Considering the motion of null particles in the spacetime of Eq.~(\ref{sigma}) and without loss of generality assuming the motion to take place along $\theta = \frac{\pi}{2}$ , we found
\begin{equation}
\dot{\Sigma}^{\,2} = E^{2} - \frac{A(\Sigma)L^{2}}{\Sigma^{2}},    
\end{equation}
where $E$ and $L$ are the conserved quantities previously defined and the dot represents differentiation with respect to some affine parameter. Redefining $U \equiv \frac{1}{\Sigma}$, differentiating with
respect to $\phi$, we can write
\begin{equation}
U''+U = 3MU^{2} -2\rho_{0}U^{3}.
\end{equation}
To solve this equation, we will employ a perturbative method, as in the previous case, treating the general-relativistic corrections as small deviations from
the Newtonian solution.
\begin{equation}
U(\phi) = U_{0}(\phi) + U_{1}(\phi) + \cdots,
\end{equation}
 Solving for the zero order, we have:
\begin{equation} U_{0}''+U_{0}=0.
\end{equation}
which leads to,
\begin{equation}\label{zeroroder1}
U_{0} = \frac{\sin\phi}{D}.
\end{equation}
where $D$ is a constant of integration and we impose the initial condition $\phi(0) = 0$ for convenience. This trajectory is a straight line in polar coordinates, as expected for the flat-spacetime trajectory. The first-order equation becomes:
\begin{equation}
U_{1}''+U_{1} = 3M U_{0}^{2} -2\rho_{0}U_{0}^{3},
\end{equation}
whose solution can be written in the following way
\begin{equation}
U_{1}=\frac{M}{D^2}\left(1+C\cos\phi+\cos^2\phi\right)+\frac{\rho_{0}}{D^{3}} \left(\frac{3}{4}\phi\cos\phi - \frac{1}{16}\sin 3\phi \right).
\end{equation}
where $C$ is an arbitrary constant. So, the general solution can be written as
\begin{equation}
U(\phi) = \frac{\sin\phi}{D}+ \frac{M}{D^2}\left(1+C\cos\phi+\cos^2\phi\right)+ \frac{\rho_{0}}{D^{3}}\left(\frac{3}{4}\phi\cos\phi - \frac{1}{16}\sin 3\phi \right) + \mathcal{O}(M^{2},M\rho_{0},\rho_{0}^{2}).
\end{equation}
Assuming that the source is located at $\Sigma \to \infty$ where $\phi \to - \delta_1$ and the observer is localized at $\Sigma \to \infty$ such that $\phi \to \pi +\delta_2$, where $|\delta_{1}|\ll 1$, $|\delta_{2}|\ll 1$, the total angle of deflection is given by $\delta = \delta_1 + \delta_2$. The light-ray deflection angle in the metric can be written as
\begin{equation}
\delta (D) = \frac{4M}{D}-\frac{3\pi\rho_{0}}{4D^{2}},    
\end{equation}
which coincides with the Reissner--Nordstr\"om result
\cite{Hu:2013eya}. We now express the deflection angle in terms of the
original radial coordinate $r$. Let $d\equiv r_{\rm min}^{(0)}$ denote
the radial coordinate of the turning point of the zeroth-order null
trajectory. Working directly in the $r$ coordinate, the zeroth-order
orbital equation can be written as
\begin{equation}
\left(\frac{dr}{d\phi}\right)_{0}^{2} = \frac{\left(r^{2}+r_{0}^{2}\right)^{2}}
{D^{2}r^{2}} \left(r^{2}+r_{0}^{2}-D^{2}\right),
\end{equation}
where $D=L/E$ is the zeroth-order impact parameter. At the turning point $r=d$, one has $\left(\frac{dr}{d\phi}\right)_{0}=0$, and therefore
\begin{equation}
D^{2}=d^{2}+r_{0}^{2},
\end{equation}
Assuming $\frac{r_{0}^{2}}{d^{2}}\ll1$, and retaining terms up to quadratic order in the bounce parameter, while keeping only the leading contributions in $M$ and $\rho_{0}$, we obtain
\begin{equation}
\delta(d) = \frac{4M}{d} -\frac{3\pi\rho_{0}}{4d^{2}} -\frac{2Mr_{0}^{2}}{d^{3}} +\frac{3\pi\rho_{0}r_{0}^{2}}{4d^{4}} +\mathcal{O} \left(M^{2},M\rho_{0},\rho_{0}^{2},r_{0}^{4}\right).
\end{equation}
The first term corresponds to the usual Schwarzschild result. Assuming a positive energy density, compatibility with the observational uncertainty $\sigma_\delta$ then requires
\begin{equation}\label{eq:deflection_constraint}
\delta(d)-\delta_{\rm Schw}(d) = \left|-\frac{3\pi\rho_{0}}{4d^{2}}+r_{0}^{2}\left(-\frac{2M}{d^{3}}+\frac{3\pi\rho_{0}}{4d^{4}}\right)\right|\leq \sigma_{\delta}.
\end{equation}
For $0<\rho_{0}<\frac{8Md}{3\pi}$, both contributions to $\delta(d)-\delta_{\rm Schw}(d)$ are negative. Therefore, Eq.~(\ref{eq:deflection_constraint}) yields the upper bound
\begin{equation}
r_{0,\rm max}(\rho_{0}) = \left[\frac{\sigma_{\delta} -\dfrac{3\pi\rho_{0}}{4d^{2}}}{\dfrac{2M}{d^{3}}-\dfrac{3\pi\rho_{0}}{4d^{4}}}\right]^{1/2}.
\end{equation}
In addition, the existence of a real-valued upper bound requires
\begin{equation}
\rho_{0} \leq \frac{4d^{2}\sigma_{\delta}}{3\pi}.
\end{equation}
At $\rho_{0}=\frac{8Md}{3\pi}$, the coefficient of the $r_{0}^{2}$ contribution vanishes at this order. Consequently, the light-deflection measurement does not constrain $r_{0}$ within the present approximation. For $\rho_{0}>\frac{8Md}{3\pi}$, the $\rho_{0}$ and $r_{0}^{2}$ contributions have opposite signs and may partially cancel.

\subsection{TIME DELAY OF LIGHT}
Considering once again the motion of null particles in the spacetime described by Eq.(\ref{sigma}), and assuming without loss of generality that the motion is confined to the equatorial plane, we obtain from the zeroth-order solution (\ref{zeroroder1}), the following relation
\begin{equation}
\Sigma^{2}d\phi^{2} = \frac{D^{2}}{\Sigma^{2}-D^{2}}d\Sigma^{2}.    
\end{equation}
Which allows us to write the following expression
\begin{equation}
\frac{dt}{d\Sigma} = \pm \sqrt{\frac{1}{A^{2}(\Sigma)} + \frac{D^{2}}{A(\Sigma)(\Sigma^{2}-D^{2})}}.    
\end{equation}
To first order in $M$ and $\rho_{0}$, $\frac{M}{D}\ll1, \frac{\rho_{0}}{D^{2}}\ll1$, we obtain
\begin{equation}
\frac{dt}{d\Sigma} \simeq \frac{\Sigma}{\sqrt{\Sigma^{2}-D^{2}}}\left[1 + M\left(\frac{2}{\Sigma} - \frac{D^{2}}{\Sigma^{3}}\right) + \rho_{0}\left(-\frac{1}{\Sigma^{2}}+ \frac{D^{2}}{2\Sigma^{4}}\right)\right].    
\end{equation}
The Shapiro time-delay consists of an emitter and a receiver located far from the massive source described by the black-bounce spacetime. Assuming that a light ray, or radar signal, is emitted from a source located at $\Sigma_E$, propagates along the zeroth-order trajectory with closest-approach parameter $D$, and is received at $\Sigma_R$, the corresponding coordinate travel time is given by
\begin{align}
t =&  t_0 +\frac{1}{c}\Bigg\{2M \left[ \ln\left( \frac{\Sigma_E+\sqrt{\Sigma_E^2-D^2}}{D} \right) + \ln\left(\frac{\Sigma_R+\sqrt{\Sigma_R^2-D^2}}{D}\right)\right]
\nonumber\\
&-M\left[\frac{\sqrt{\Sigma_E^2-D^2}}{\Sigma_E} + \frac{\sqrt{\Sigma_R^2-D^2}}{\Sigma_R}\right] 
+\rho_0 \Bigg[\frac{1}{4}\left(\frac{\sqrt{\Sigma_E^2-D^2}}{\Sigma_E^2} + \frac{\sqrt{\Sigma_R^2-D^2}}{\Sigma_R^2}
\right) \nonumber\\
&-\frac{3}{4D} \left(\arccos\left(\frac{D}{\Sigma_E}\right)
+ \arccos\left(\frac{D}{\Sigma_R}\right)\right)\Bigg]\Bigg\},
\end{align}
where,
\begin{equation}
t_0 = \frac{1}{c}\left(\sqrt{\Sigma_E^2-D^2} + \sqrt{\Sigma_R^2-D^2}\right),    
\end{equation}
represents the travel time in flat spacetime. Considering that for Solar-System applications, the closest-approach distance is much smaller than the distances to the emitter and receiver, the total round-trip travel time then becomes
\begin{equation}
\delta T \equiv T-T_{0} = \frac{4M}{c} \left[\ln\left(\frac{4\Sigma_{E}\Sigma_{R}}{D^{2}}\right)-1\right]-\frac{3\pi\rho_{0}}{2cD}. 
\end{equation}
with $T_0 = 2t_0$, which is consistent with the Reissner-Nordstrom result \cite{Junior:2023nku}. Rewriting this expression in terms of the original radial coordinate and assuming $d\ll r_E,r_R$, together with $\frac{r_0^2}{d^2}\ll1$,$\frac{r_0^2}{r_E^2}\ll1$,$\frac{r_0^2}{r_R^2}\ll1$, we obtain
\begin{equation}
\delta T \simeq \frac{4M}{c} \left[\ln\left(\frac{4r_Er_R}{d^2}\right)-1 \right] -\frac{3\pi\rho_0}{2cd}
+ \frac{r_0^2}{c}\left(-\frac{4M}{d^2} +\frac{3\pi\rho_0}{4d^3} \right) +\mathcal{O}(r_0^4).
\end{equation}
The first term corresponds to the Schwarzschild prediction. Assuming a positive energy density, $\rho_0>0$, compatibility with the observational uncertainty, $\sigma_{\Delta T}$, requires
\begin{equation}
\left|-\frac{3\pi\rho_0}{2d} + r_0^2 \left( -\frac{4M}{d^2} +\frac{3\pi\rho_0}{4d^3}\right) \right| \leq c\sigma_{\Delta T}.
\label{eq:time_delay_constraint}
\end{equation}
For $0<\rho_0<\frac{16Md}{3\pi}$, both corrections are negative and no cancellation occurs. In this regime,
Eq.~(\ref{eq:time_delay_constraint}) yields the upper bound
\begin{equation}
r_{0,\rm max}(\rho_0) = \left(\frac{c\sigma_{\Delta T}-\dfrac{3\pi\rho_0}{2d}}{\dfrac{4M}{d^2}-\dfrac{3\pi\rho_0}{4d^3}}\right)^{1/2}.
\end{equation}
A real-valued bound requires
\begin{equation}
\rho_0 \leq \frac{2cd\,\sigma_{\Delta T}}{3\pi}.
\end{equation}
At $\rho_0=\frac{16Md}{3\pi}$, the correction proportional to $r_0^2$ vanishes at this order. For $\rho_0>\frac{16Md}{3\pi}$, the two corrections have opposite signs and may partially cancel.

\section{SHADOW RADIUS}
Following the analysis of Ref.~\cite{Alencar:2026qeb}, the shadow radius
can be indirectly constrained by combining the angular size of the observed image with independent measurements of the mass-to-distance ratio. Using the Keck \cite{Do:2019txf} and VLTI measurements \cite{GRAVITY:2020gka}, the corresponding observational bounds are
\begin{equation}
4.55 \lesssim \frac{r_S}{M} \lesssim 5.22
\qquad (1\sigma), \qquad 4.21 \lesssim \frac{r_S}{M} \lesssim 5.56 \qquad (2\sigma).
\label{shadowbounds}
\end{equation}
To determine the theoretical shadow radius, we first consider the unstable circular photon orbit. For null geodesics confined to the equatorial plane, $\theta=\frac{\pi}{2}$, the radial equation can be written as
\begin{equation}
\dot{\Sigma}^{\,2} = E^2\left[1-b^2\frac{A(\Sigma)}{\Sigma^2}\right],
\label{nullradial}
\end{equation}
where, $b\equiv\frac{L}{E}$ is the impact parameter. It is convenient to introduce the null effective potential
\begin{equation}
V(\Sigma) = \frac{A(\Sigma)}{\Sigma^2}.
\label{photonpotential}
\end{equation}
An unstable circular photon orbit at $\Sigma=\Sigma_{\rm ph}$ satisfies
\begin{equation}
V(\Sigma_{\rm ph}) = \frac{1}{b_c^2}, \qquad \left. \frac{dV}{d\Sigma} \right|_{\Sigma=\Sigma_{\rm ph}} =0, \qquad \left. \frac{d^2V}{d\Sigma^2} \right|_{\Sigma=\Sigma_{\rm ph}} <0,
\label{photonconditionssigma}
\end{equation}
where $b_c$ denotes the critical impact parameter. The outer unstable branch is therefore
\begin{equation}
\Sigma_{\rm ph} = \frac{3M+\sqrt{9M^2-8\rho_0}}{2}.
\label{sigmaph}
\end{equation}
Accordingly, the existence of a real photon-sphere solution requires
\begin{equation}
9M^2-8\rho_0\geq0.
\end{equation}
Following Ref.~\cite{Alencar:2026qeb}, the angular radius $\alpha$ measured by a static observer located at $\Sigma=\Sigma_O$ is related to the impact parameter by
\begin{equation}
\sin^2\alpha = \frac{b^2 A(\Sigma_O)}{\Sigma_O^2}.
\label{sinalpha}
\end{equation}
For the critical null geodesic, Eq.~(\ref{photonconditionssigma}) gives
\begin{equation}
b_c^2 = \frac{\Sigma_{\rm ph}^2}{A(\Sigma_{\rm ph})}.
\label{criticalimpact}
\end{equation}
Hence, the shadow radius measured by a static observer is
\begin{equation}
r_S = \Sigma_{\rm ph}\sqrt{\frac{A(\Sigma_O)}{A(\Sigma_{\rm ph})}}.
\label{shadowfinite}
\end{equation}
For an asymptotically distant observer,
$A(\Sigma_O)\rightarrow1$, and therefore
\begin{equation}
r_S = \frac{\Sigma_{\rm ph}}{\sqrt{A(\Sigma_{\rm ph})}}.
\label{shadowsigma}
\end{equation}
Introducing the dimensionless quantities
\begin{equation}
\bar{\rho}\equiv\frac{\rho_0}{M^2},
\qquad
x_{\rm ph}\equiv\frac{\Sigma_{\rm ph}}{M} = \frac{3+\sqrt{9-8\bar{\rho}}}{2},
\end{equation}
the theoretical shadow radius becomes
\begin{equation}
\frac{r_S}{M} = \frac{x_{\rm ph}}{\sqrt{1-\dfrac{2}{x_{\rm ph}}+\dfrac{\bar{\rho}}{x_{\rm ph}^2}}}.
\label{shadowrho}
\end{equation}
For positive $\rho_0$, comparison with the observational intervals in
Eq.~(\ref{shadowbounds}) gives approximately
\begin{equation}
0\leq\frac{\rho_0}{M^2}\lesssim0.637
\qquad (1\sigma),
\end{equation}
and
\begin{equation}
0\leq\frac{\rho_0}{M^2}\lesssim0.882 \qquad (2\sigma).
\end{equation}
At this point, it is important to clarify the role of the original radial coordinate $r$ and, in particular, why the transformation from the areal radius $\Sigma$ to $r$ has a different consequence here from that found in the classical tests. In the perihelion, light-deflection, and time-delay analyses, the relevant radial scales---such as the pericenter and apocenter, the closest-approach distance, and the emitter and receiver positions---are independent orbital or boundary parameters. When their corresponding areal
radii are rewritten according to
\begin{equation}
\Sigma^2=r^2+r_0^2,
\end{equation}
explicit corrections depending on $r_0$ remain in the corresponding observables. The situation is different for the shadow. The photon-sphere radius is not an independently specified radial scale. Instead, its areal radius $\Sigma_{\rm ph}$ is dynamically fixed by the circular null-orbit condition in Eq.~(\ref{photonconditionssigma}). For the outer branch considered here, Eq.~(\ref{sigmaph}) shows that $\Sigma_{\rm ph}$ depends only on $M$ and
$\rho_0$. Reexpressing its position in terms of the original coordinate therefore gives
\begin{equation}
\Sigma_{\rm ph}^2 = r_{\rm ph}^2+r_0^2,
\end{equation}
or
\begin{equation}
r_{\rm ph} = \sqrt{\left(\frac{3M+\sqrt{9M^2-8\rho_0}}{2}\right)^2-r_0^2}.
\label{rphoriginal}
\end{equation}
Thus, $r_{\rm ph}$ is not independent of $r_0$; rather, it changes with the bounce parameter so as to preserve the areal photon-sphere radius fixed by the circular-orbit condition. This can also be seen by writing the asymptotic shadow radius directly in terms of the original radial coordinate,
\begin{equation}
r_S = \frac{\sqrt{r_{\rm ph}^2+r_0^2}}{\sqrt{1-\dfrac{2M}{\sqrt{r_{\rm ph}^2+r_0^2}}
+\dfrac{\rho_0}{r_{\rm ph}^2+r_0^2}}}.
\label{shadowr}
\end{equation}
Although Eq.~(\ref{shadowr}) appears to contain an explicit dependence on $r_0$, the circular photon-orbit condition imposes
\begin{equation}
r_{\rm ph}^2+r_0^2 = \Sigma_{\rm ph}^2 = \left(\frac{3M+\sqrt{9M^2-8\rho_0}}{2}\right)^2.
\end{equation}
Consequently, the $r_0$ dependence cancels identically from the critical impact parameter and hence from the asymptotic shadow radius. Therefore, in contrast to the classical tests, rewriting the shadow observable in terms of the original radial coordinate does not generate an
independent $r_0$-dependent correction. The coordinate location of the photon sphere depends explicitly on the bounce parameter, whereas its areal radius and the associated critical impact parameter do not. The parameter $r_0$ instead determines whether the outer photon-sphere branch can be realized in the black-bounce geometry. Requiring $r_{\rm ph}^2>0$ gives
\begin{equation}
\frac{r_0}{M}<\frac{3+\sqrt{9-8\rho_0/M^2}}{2}.
\label{photonexistence}
\end{equation}
Hence, the shadow-radius measurement directly constrains $\rho_0$, while $r_0$ enters through the existence condition of the outer unstable photon orbit rather than through an independent modification of the asymptotic shadow size.

\section{Results}
\begin{figure}[t]
\includegraphics[height=5cm]{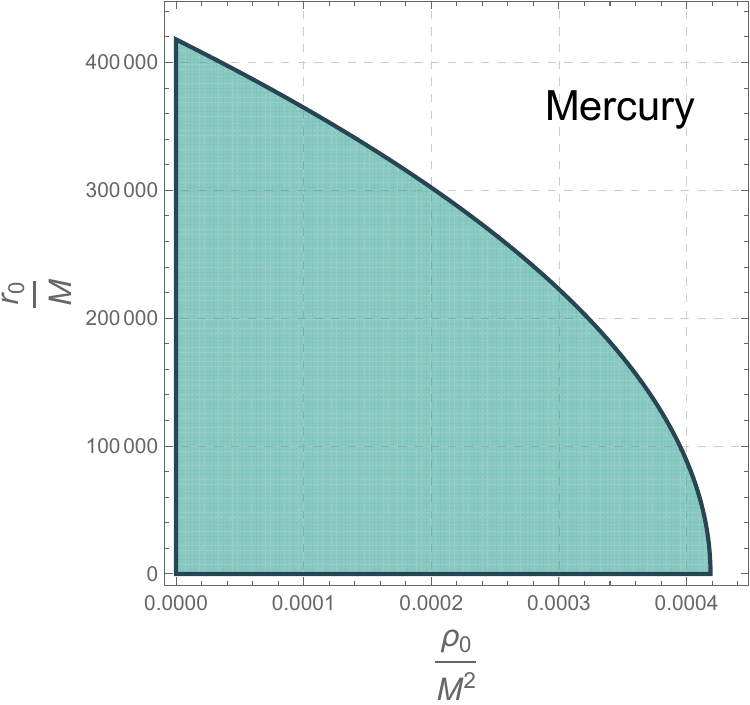}
\includegraphics[height=5cm]{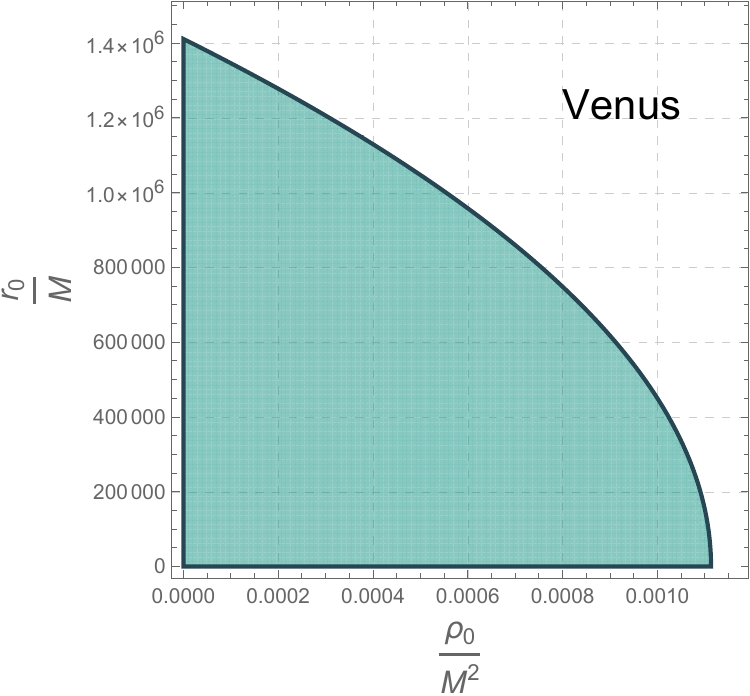}
\includegraphics[height=5cm]{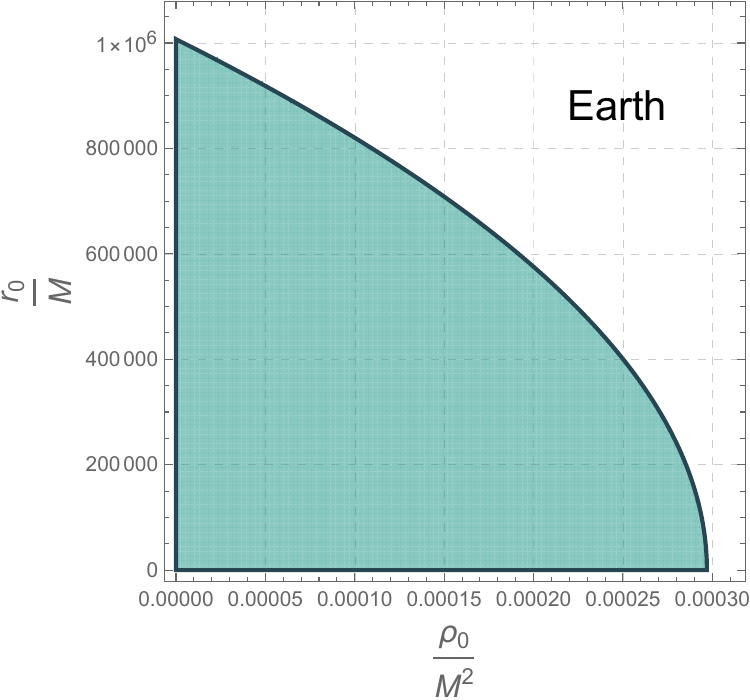}\\
(1a) \hspace{5cm}(1b) \hspace{5cm} (1c)\\
\includegraphics[height=5cm]{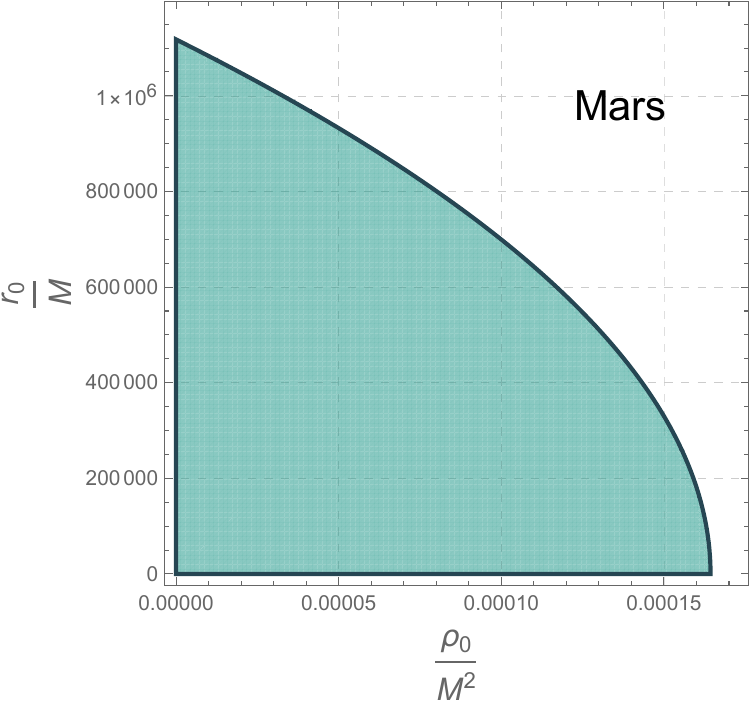}
\includegraphics[height=5cm]{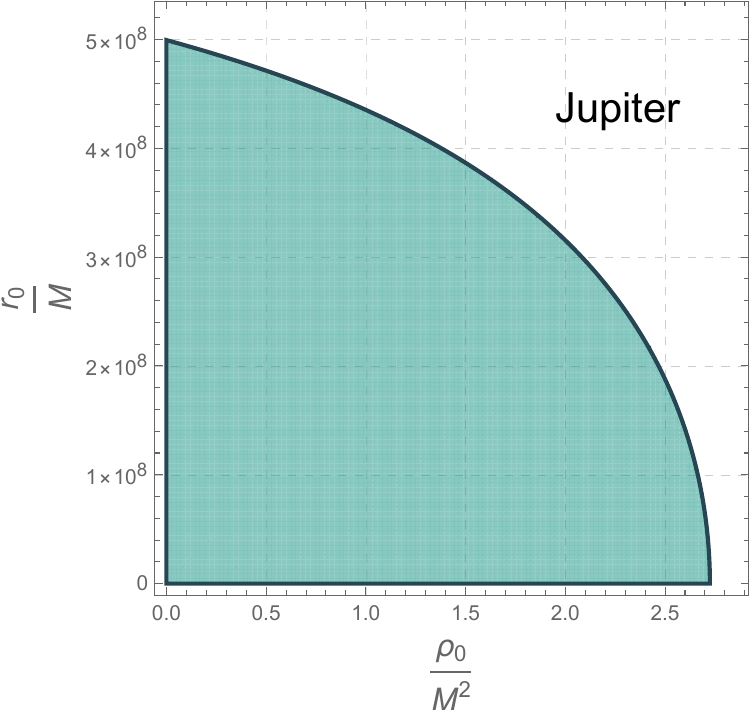}
\includegraphics[height=5cm]{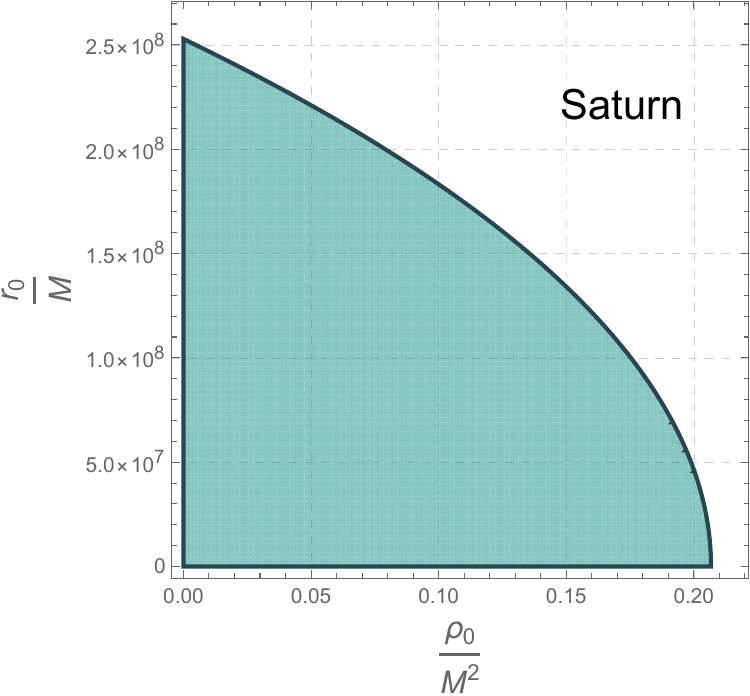}\\
(1d) \hspace{5cm}(1e) \hspace{5cm} (1f)\\
\includegraphics[height=5cm]{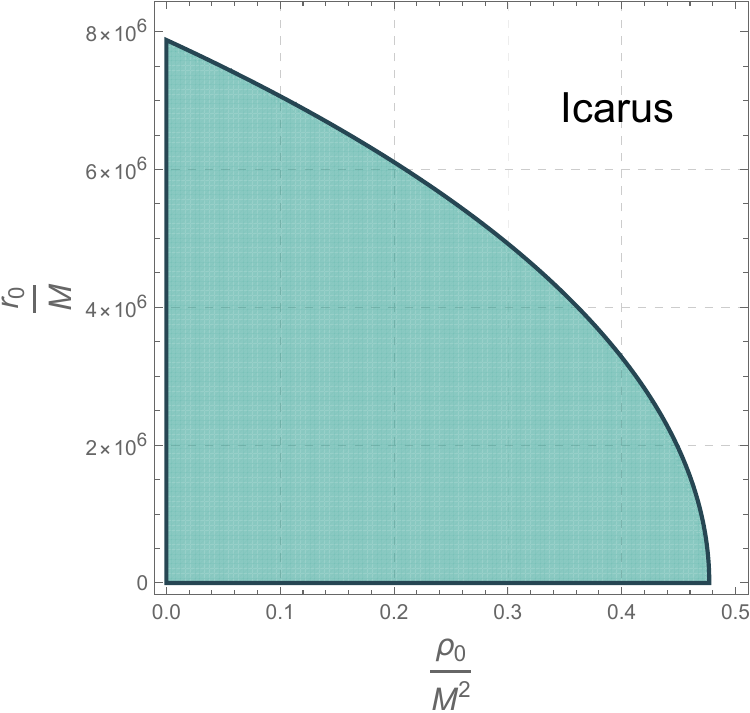}\\
(1g)\\
\caption{Parameter space $(r_0,\rho_0)$ for the perihelion precession of several planets and the asteroid Icarus. We adopt the orbital parameters from the NASA Planetary Fact Sheet and the JPL Small-Body Database, while the observational uncertainties in the perihelion advances of the planets are taken from Ref.~\cite{Pitjeva:2013xxa}, and that of Icarus from Refs. \cite{Shapiro:1968zza,Shapiro:1971iv}.\label{Figure1}}
\end{figure}

We now translate the observational uncertainties discussed in the previous sections into constraints on the parameters of the symmetric black-bounce. Throughout this section, we introduce the dimensionless quantities
\begin{equation}
\widehat{r}_{\rm ph}\equiv\frac{r_{\rm ph}}{M}, \qquad \widehat{\rho}\equiv \frac{\rho_0}{M^2}, \qquad \widehat r_0\equiv \frac{r_0}{M}.
\end{equation}
For the Solar-System tests we set $M=\frac{G M_\odot}{c^2}$ and restrict the analysis to $\rho_0>0$. In each panel below, the shaded region represents the combinations of $(\widehat{\rho},\widehat r_0)$ that satisfy the corresponding observational constraint, whereas the solid boundary corresponds to saturation of the experimental uncertainty. Since the weak-field expressions were obtained by expanding in powers of $r_0^2$ relative to the characteristic orbital or impact-parameter scale, only the portion of the parameter space satisfying the corresponding perturbative conditions should be regarded as quantitatively reasonable. We first consider the perihelion advance. Following Ref.~\cite{Casana:2017jkc}, we use the measurements for Mercury, Venus, Earth, Mars, Jupiter, Saturn, and the asteroid Icarus. The corresponding allowed regions in the $(\widehat{\rho},\widehat r_0)$ plane are shown in Fig. \ref{Figure1}. The different scales displayed in Fig. \ref{Figure1} show that the perihelion data constrain the matter parameter $\rho_0$ considerably more efficiently than the bounce parameter $r_0$. In particular, the inner-planet measurements lead to much smaller allowed values of $\widehat{\rho}$ than those obtained from Jupiter and Saturn. On the other hand, the upper limits on $\widehat r_0$ remain very large, reflecting the weak sensitivity of planetary orbits to the minimum areal radius.

\begin{figure}[t]
\includegraphics[height=5.1cm]{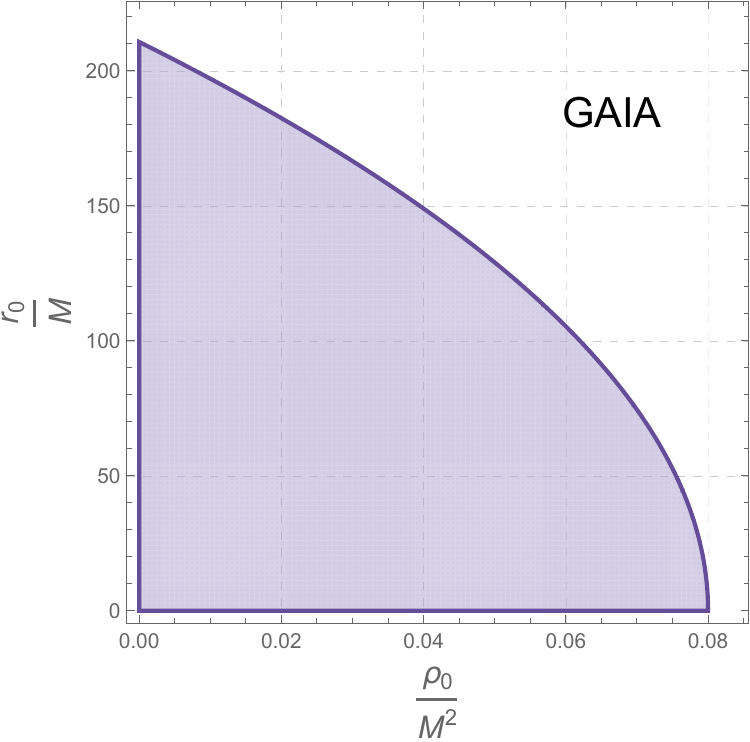}
\includegraphics[height=5.1cm]{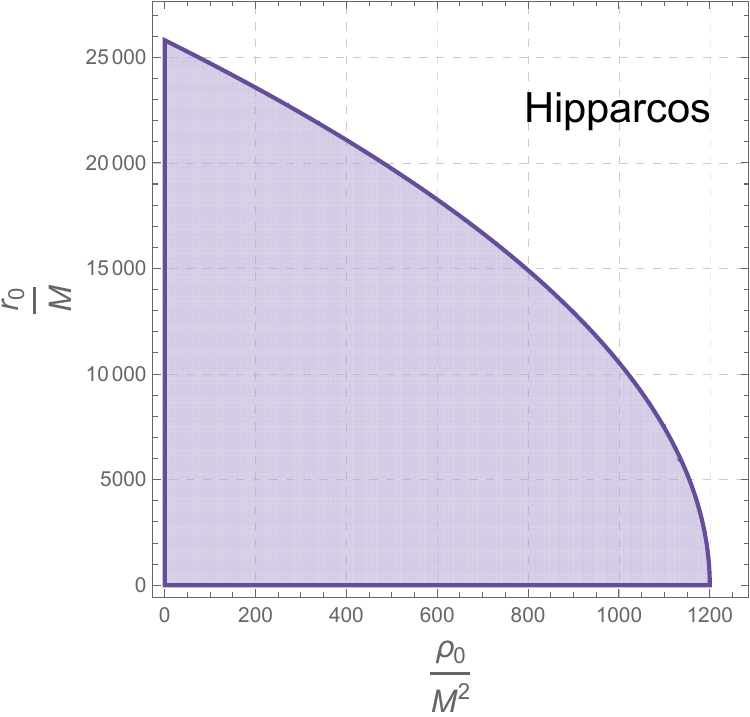}
\includegraphics[height=5.1cm]{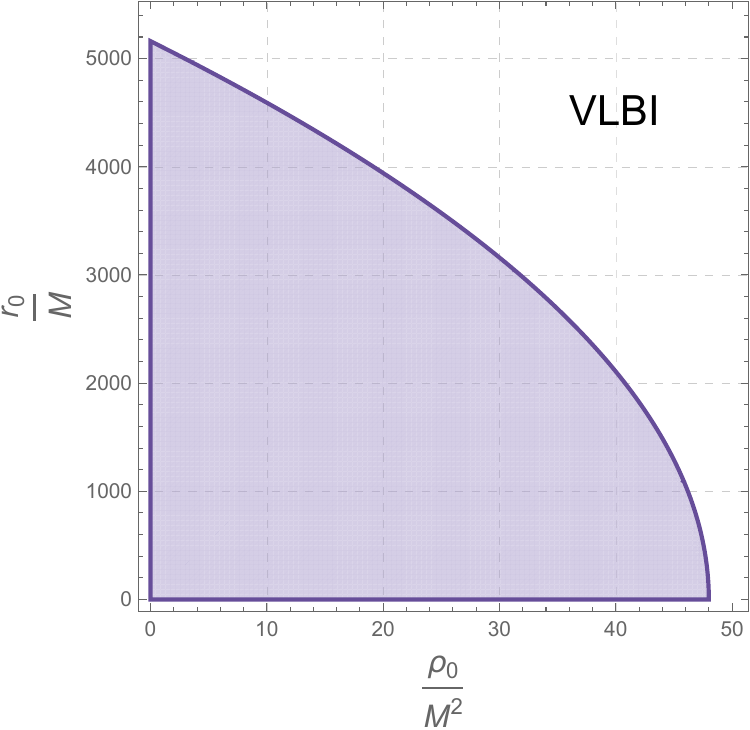}\\
(2a) \hspace{4 cm}(2b) \hspace{4cm} (2c)\\ 
\includegraphics[height=5.1cm]{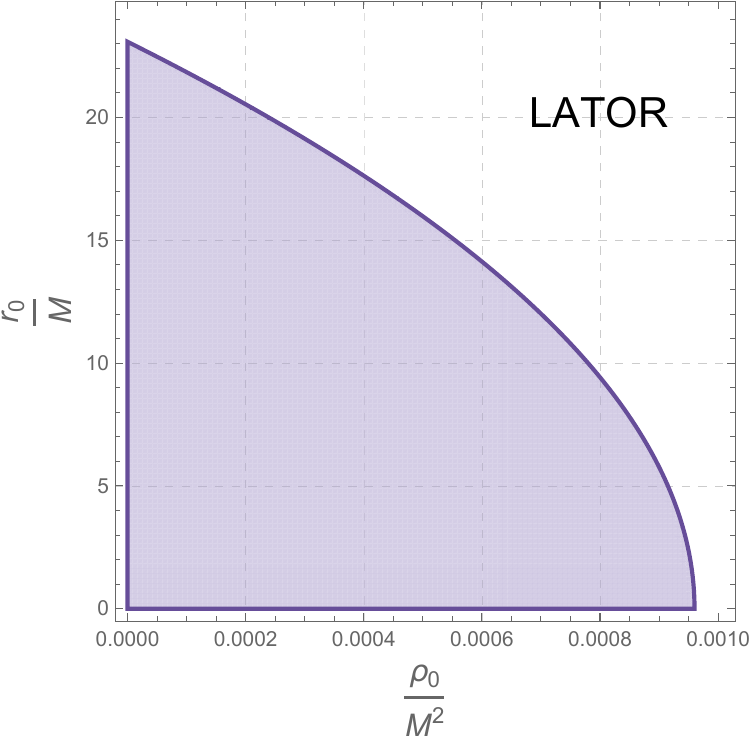}
\includegraphics[height=5.1cm]{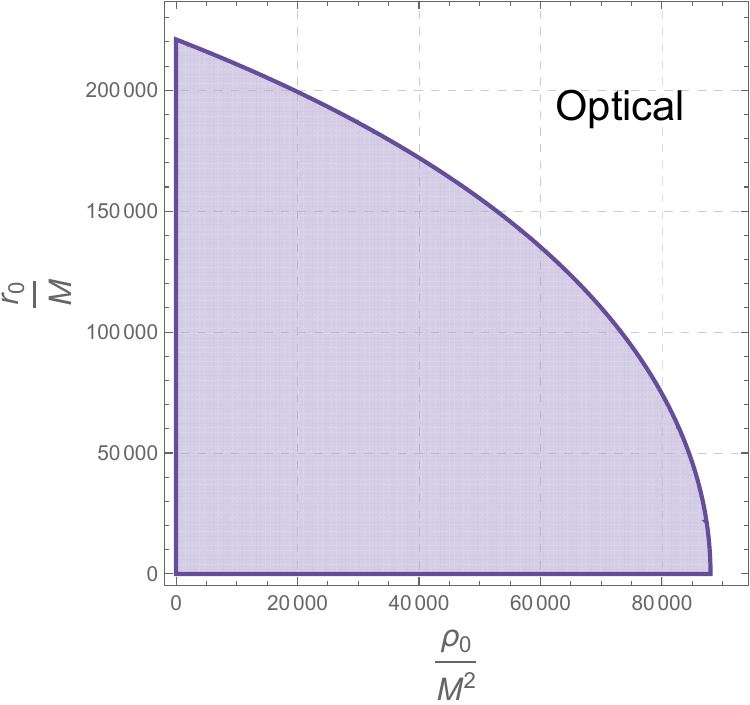}\\
(2d) \hspace{4 cm} (2e)\\
\caption{Parameter space $(r_0,\rho_0)$ allowed by light-deflection measurements. We consider a light ray grazing the
solar limb, with $d\simeq R_\odot$. The observational sensitivities adopted
for LATOR, GAIA, VLBI, Hipparcos, and ground-based optical measurements
are taken from Refs.~\cite{Plowman:2005fb,Vecchiato:2003av,Lambert:2009xy,Froeschle:1997,TexasMauritanianEclipse:1976nev}.\label{Figure2}}
\end{figure}

For the light-deflection test, we consider a ray grazing the solar vicinities and use the sensitivities associated with GAIA, Hipparcos, VLBI, LATOR, and ground-based optical measurements. The corresponding parameter regions are shown in Fig. \ref{Figure2}. As expected, the size of the allowed region decreases as the angular resolution improves. Among the configurations considered here, LATOR provides the strongest constraint on both parameters. This is particularly evident for the bounce parameter: the allowed values of $\frac{r_0}{M}$ obtained from LATOR are several orders of magnitude smaller than those associated with the less precise optical measurements. Nevertheless, the same hierarchy observed in the perihelion analysis persists: the matter contribution is more efficiently constrained than the
minimum-radius parameter. The latter enters the weak-field deflection only through corrections suppressed by powers of $ \frac{r_0}{d}$, which explains the comparatively weak sensitivity to the bounce scale.

\begin{figure}[t]
\includegraphics[height=5.1cm]{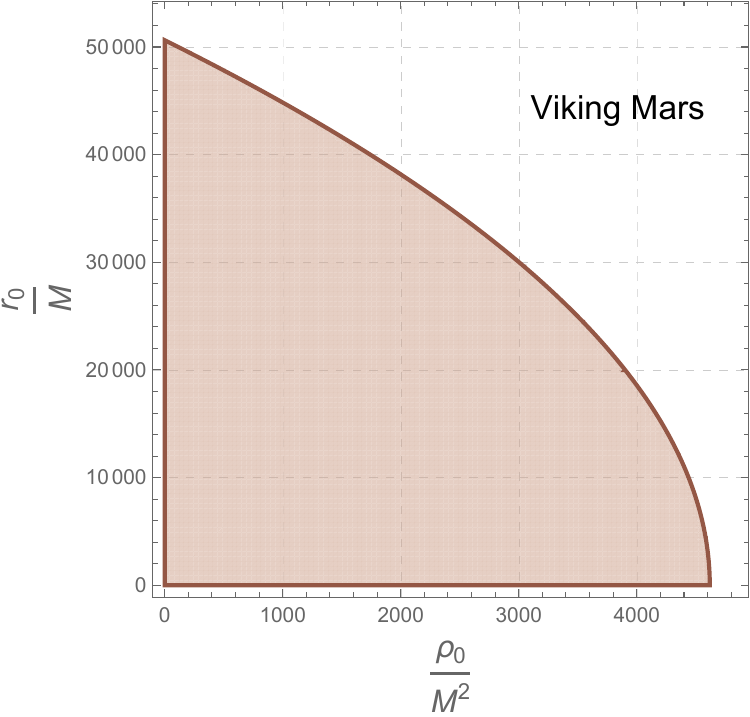}
\includegraphics[height=5.1cm]{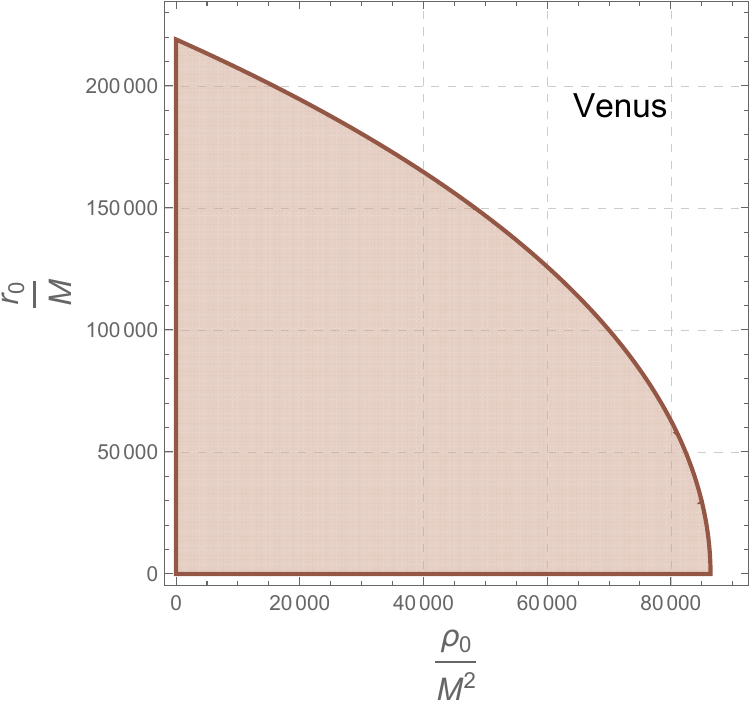}
\includegraphics[height=5.1cm]{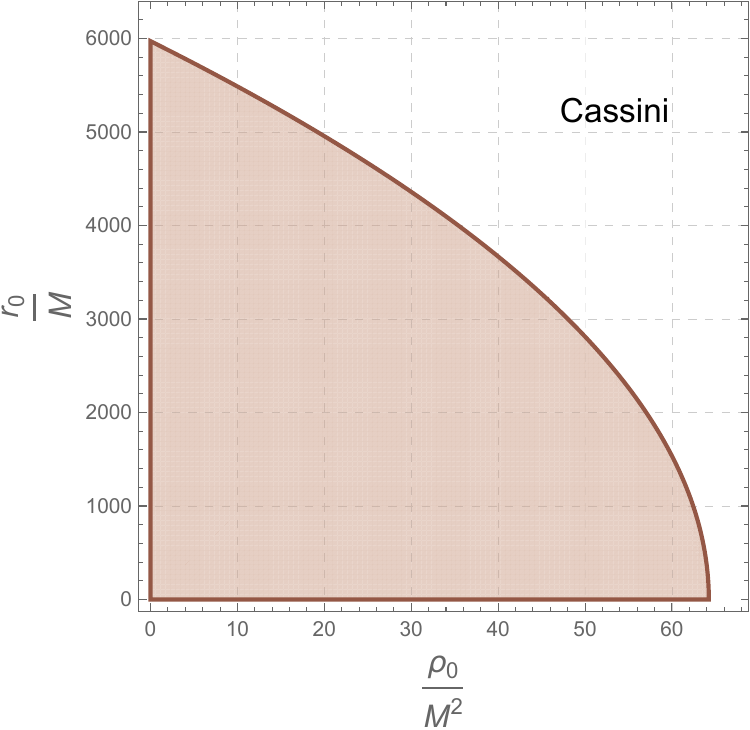}\\
(3a) \hspace{5cm}(3b) \hspace{5cm} (3c)\\
\caption{Parameter space $(r_0,\rho_0)$ allowed by Shapiro time-delay measurements. We consider radar signals passing close to the solar limb, with $d\simeq R_\odot$ and $r_E\simeq 1\,\mathrm{AU}$.
We adopt fractional observational accuracies of $2\%$, $0.1\%$, and $1.2\times10^{-5}$ for the Venus radar-ranging, Viking Mars, and Cassini
experiments, respectively, following Refs.~\cite{Shapiro:1971iv,Reasenberg:1979ey,Bertotti:2003rm}.\label{Figure3}}
\end{figure}

For the Shapiro time-delay test, we consider the Venus radar-ranging experiment, the Viking Mars experiment, and the Cassini measurement. We take $r_E\simeq1\,{\rm AU}$ and approximate the receiver distances by $r_R\simeq0.72\,{\rm AU}$ and $1.52\,{\rm AU}$ for Venus and Mars, respectively, using their mean heliocentric orbital distances \cite{NASAPlanetaryFactSheet}. For the Cassini configuration, we take $r_R\simeq7.43\,{\rm AU}$, as inferred from the geocentric distance reported in Ref.~\cite{Bertotti:2003rm}. We further adopt $d\simeq R_\odot$ as a common closest-approach scale for the Solar-System comparison. We adopt fractional observational accuracies of $2\%$, $0.1\%$, and $1.2\times10^{-5}$ for the Venus radar-ranging, Viking Mars, and Cassini experiments, respectively, following
Refs.~\cite{Reasenberg:1979ey,Bertotti:2003rm}. The resulting allowed regions are displayed in Fig. \ref{Figure3}. The progressive reduction of the parameter space from Venus to Viking Mars and finally to Cassini is mainly driven by the improvement in experimental precision. The GR excess delays for these configurations are all of the same order, $\Delta T_{\rm GR}\sim10^{-4}\,{\rm s}$, whereas the fractional uncertainty decreases by several orders of magnitude. Consequently, Cassini provides the strongest time-delay constraint among the measurements considered here. As in the previous weak-field tests, the allowed values of $\rho_0$ are
substantially more restricted than those of $r_0$. The relatively large values of $\frac{r_0}{M}$ that remain compatible with the data should therefore be interpreted as evidence of the weak sensitivity of weak-field observables to the minimum-radius scale, rather than as evidence favoring a large bounce.

\begin{figure}[t]
\includegraphics[height=7cm]{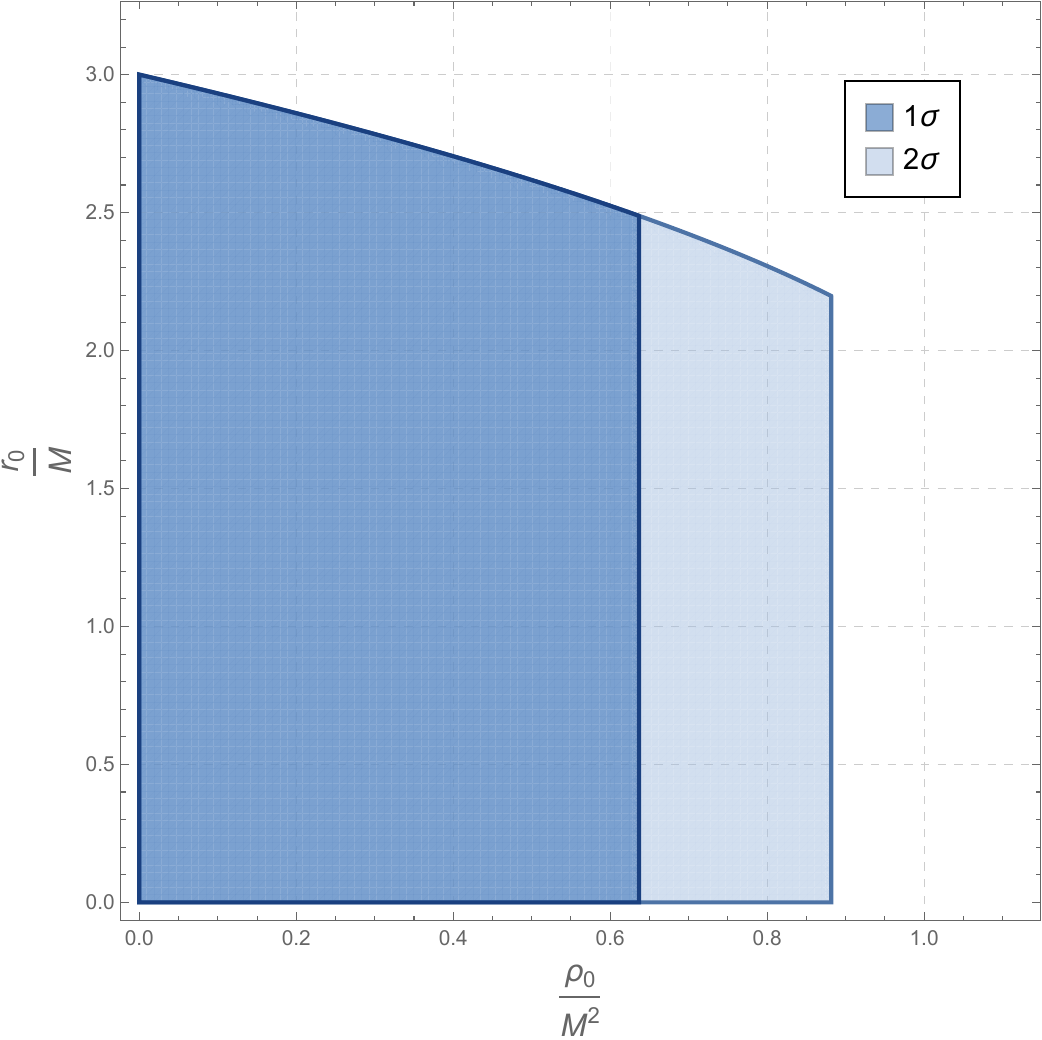}
\caption{Parameter space ($r_0,\rho_0$) for the outer photon-sphere branch compatible with the  1$\sigma$ and 2$\sigma$ shadow-radius intervals inferred from the Keck and VLTI measurements.The curved upper boundary corresponds to the existence condition of the
oute unstable photon orbit.\label{Figure4}}
\end{figure}

We finally consider the strong-field constraints associated with the black-hole shadow. For the outer photon-sphere branch, the coordinate position $r_{\rm ph}$ depends explicitly on both $\rho_0$ and $r_0$. However, the circular null-orbit condition fixes the combination
$r_{\rm ph}^2+r_0^2=\Sigma_{\rm ph}^2$ independently of $r_0$. Consequently,
the critical impact parameter, and hence the asymptotic shadow radius, depend only on the matter parameter $\rho_0$. The Keck and VLTI measurements therefore provide a direct constraint on $\rho_0$, while the bounce parameter enters through the existence condition of the outer unstable photon orbit,
\begin{equation}
\frac{r_0}{M} < \frac{3+\sqrt{9-8\rho_0/M^2}}{2}.
\end{equation}
Accordingly, the regions shown in  Fig.~\ref{Figure4} represent the intersection
between the observational shadow-radius bounds and the parameter domain in which the outer photon-sphere branch exists. The $1\sigma$ region is more restrictive than the corresponding $2\sigma$ region because of the tighter observational interval on the shadow radius. The curved upper boundary, on the other hand, should be interpreted as the limiting condition for the
existence of the outer photon sphere rather than as a direct observational upper bound on $r_0$. In contrast to the Solar-System tests, the shadow probes the strong-field geometry at scales of order $M$. Nevertheless, for the outer photon-sphere branch considered here, the shadow size itself does not provide an independent observational constraint on the bounce parameter.

\section{Summary and conclusion}\label{secV}
In this work, we derived observational constraints on the parameters of the symmetric black-bounce geometry using the perihelion advance, light deflection, time-delay, and black-hole-shadow observables. Within the weak-field analysis, constraints on both $\rho_0$ and $r_0$ are obtained when the observables are expressed in terms of the original radial coordinate $r$. While the matter parameter $\rho_0$ is relatively well constrained, the corresponding upper bounds on $r_0$ remain very large, reflecting the weak sensitivity of Solar-System observables to the minimum-radius scale.

The strong-field analysis reveals a different behavior. Although the coordinate location of the outer photon sphere depends explicitly on both
$\rho_0$ and $r_0$, the circular null-orbit condition fixes the combination $r_{\rm ph}^2+r_0^2=\Sigma_{\rm ph}^2$ independently of $r_0$. Consequently, the critical impact parameter, and hence the asymptotic shadow radius, depend only on $\rho_0$. The shadow size therefore provides a direct observational
constraint on the matter parameter, whereas $r_0$ determines the parameter domain in which the outer unstable photon orbit exists. The shadow
measurements thus complement the weak-field tests by probing the photon-sphere structure of the geometry, although the resulting constraint on $\rho_0$ is weaker than those obtained from Solar-System observations.

\begin{acknowledgments}
A. C. L. Santos would like to thank CAPES for financial support under Grant No. 88887.822058/2023-00. M. S. Melo acknowledges financial support from the Coordenação de Aperfeiçoamento de Pessoal de Nível Superior (CAPES) under Grant No. 
88887.200103/2025-00. R. V. Maluf acknowledges financial support from the Conselho Nacional de Desenvolvimento Científico e Tecnológico (CNPq) under Grant No. 311393/2025-0 (PQ).
\end{acknowledgments}

\end{document}